\documentclass[10pt,conference]{IEEEtran} 
\usepackage{array}

\usepackage{mdwmath}
\usepackage{mdwtab}

\usepackage{flushend}           

\usepackage{url}
\usepackage{times}
\usepackage[pdftex]{graphicx}	

\usepackage[numbers,sort&compress]{natbib}

\usepackage{versions}

\usepackage{hyperref}
\usepackage[capitalize]{cleveref}

\usepackage[caption=false,font=footnotesize]{subfig}

\usepackage{paralist}           

\usepackage{algorithmic}       
\usepackage{listings}           
\usepackage{moreverb}           

\usepackage{fancyvrb}           

\usepackage{eurosym}
\usepackage{units}              

\usepackage{booktabs}
\usepackage{tikz}               
\usepackage{ctable}             



\usepackage{rotating}           

\usepackage{multirow}

\newcommand{\PreserveBackslash}[1]{\let\temp=\\#1\let\\=\temp}

\usepackage[normalem]{ulem}     
\usepackage{etoolbox}           

\usepackage{footnote}
\usepackage{xcolor}
\usepackage[]{draftwatermark}
\SetWatermarkText{PREPRINT}
\SetWatermarkColor[gray]{0.9}
\SetWatermarkAngle{45}
\SetWatermarkScale{1}
 
\usepackage{xspace}             

\usepackage{fancyhdr}
 
\newtoggle{anonymous}           
\toggletrue{anonymous}          

\usepackage{xspace}             

\definecolor{pink}{rgb}{1,.4,.4}
\definecolor{red}{rgb}{1, .2, .1}

\newcommand{\Helena}{HELENA\xspace}

\newcommand{\GSD}{GSD\xspace}

\newcommand{\GSDlong}{Global Software Development\xspace}

\iftoggle{anonymous}{%

}{%

}

\begin{document}

\pagestyle{fancy}
\fancyhead[C]{\footnotesize{This is the author's version of the work. It is posted here by permission of ACM/IEEE for your personal use. \\ Not for redistribution. The definitive version was published in the ACM/IEEE International Symposium on \\ Empirical Software Engineering and Measurement (ESEM)} }
\renewcommand{\headrulewidth}{0pt} 
\SetWatermarkText{PREPRINT}
\SetWatermarkColor[gray]{0.9}
\SetWatermarkAngle{45}
\SetWatermarkScale{1}

\title{Plan-Driven Approaches Are Alive and Kicking in Agile Global Software Development}
\author{
Marcelo Marinho$^{a,b}$, John Noll$^{c,b}$, Ita Richardson$^{b}$, Sarah Beecham$^{b}$\\

\normalsize $^{a}$ Federal Rural University of Pernambuco (UFRPE)  Department of Computer Science (DC) Recife, PE, Brazil\\
Email: {marcelo.marinho@ufrpe.br}\\

\normalsize $^{c}$ University of East London
University Way, London, E16 2RD, UK\\
Email: {j.noll@uel.ac.uk}\\

\normalsize $^{b}$ Lero, the Irish Software Research Centre
University of Limerick, Ireland\\
Email: {ita.richardson,sarah.beecham@lero.ie}\\

}

\IEEEoverridecommandlockouts 

\maketitle

\begin{abstract}
\noindent
\newline\textit{Background:} Agile methods are no longer restricted to small projects and co-located teams. The last decade has seen the spread of agile into large scale, distributed and regulated domains. Many case studies show successful agile adoption in GSD, however, taken as a whole, it remains unclear how widespread this trend is, and what form the agile adoption takes in a global software development (GSD) setting. \newline
\textit{Aims:} Our objective is to gain a deeper understanding of how organisations adopt agile development methods in distributed settings. Specifically we aim to plot the current development process landscape in GSD. \newline
\textit{Method:} We analyse industrial survey data from 33 different countries collected as part of the \Helena project that explored the wider use of hybrid development approaches in software development. We extract and analyse the results of 263 surveys completed by participants involved in globally distributed projects. \newline
\textit{Results:} In our sample, 72\% of globally distributed projects implement  a mix of both agile and traditional approaches (termed `hybrid'). 25\% of GSD organisations are predominantly agile,  with only very few (5\%)  opting for traditional approaches. GSD projects that used only agile methods tended to be very large. \newline
\textit{Conclusions:} Globally Distributed Software Development (and project size) is not a barrier to adopting agile practices. Yet, to facilitate project coordination and general project management, many adopt traditional approaches, resulting in a hybrid approach that follows defined rules. \newline

\end{abstract}


\section{Introduction}\label{sec:intro}
Agile methods have become common in software development organizations around the world. Initially, the methods were used for the development of small, co-located projects. However, in recent years many large organizations have made the transition from traditional, plan-driven waterfall-type methods towards agile methods \cite{boehm2005management}. According to the largest reoccurring survey on agile adoption, the State of Agile Survey \cite{versionOne}, 86\% of respondents had at least some distributed teams adopting agile practices. While this survey is not scientific, it indicates that a significant number of large global organizations are adopting agile methods.

VersionOne \cite{agileuprising} identified two reasons as to why companies use agile methods: some  believe in the values and principles of the manifesto, and others  see agile as best practice. However, Boehm and Turner \cite{boehm2005management} suggest that projects should find a `sweet spot' combining a mixture of traditional and agile methods. According to their certainty/ambiguity puzzle, business management still demands accurate and complete long-term estimates of projects and tasks, whereas development teams have to deal with uncertainty during the project \cite{marinho2018global}. Previous research on development methods advise that methods need to be adapted to the work context \cite{vinekar2006can,akbar2015short}. Hence, software development has to balance the need for planning and controlling and, at the same time, for flexibility and speed \cite{kuhrmann2017hybrid,theocharis2015water}. 

The many development approaches adopt different philosophies (e.g. agile, traditional) and specific features (e.g. plan-driven, iterative-incremental). Such approaches comprise either practices or comprehensive process frameworks. Furthermore, software development has become vital in every industry sector and must adhere to domain-specific standards, norms, and regulations. In this regard, practitioners have begun developing so-called hybrid development approaches. 

Acknowledged as one of the trends of the 21st century, globalization significantly changed many industries, including and, in particular, software development. Many companies foster global software development (GSD) to benefit from cheaper, faster, and better software development \cite{vsmite2010empirical}. However, GSD has traditionally followed a plan-driven approach, where tasks are allocated according to where they appear in the software development lifecycle \cite{estler2014agile,ylikotila2011collaboration}. The belief that agile methods, which were mainly used for small projects and co-located teams, cannot be used in GSD no longer stands \cite{noll2016agile,ramesh2006can}. Agile methods tend to rely on informal processes and regular face-to-face communication to facilitate coordination, whereas distributed software development relies on formal mechanisms \cite {noll2016agile}.

In this paper, we aim to highlight the extent to which GSD projects are adopting hybrid approaches. We take a deeper look at those few organizations that do not avail of plan driven approaches to manage their software development. We also uncover what drives organizations to change, adapt or merge agile practices. This study seeks to determine whether it is possible to adopt agility in GSD projects without the support of traditional plan-driven approaches. The paper is based on the \Helena Survey~\cite{helenastage2}, an international study on the use of hybrid development approaches in software development around the world.

This paper is organized as follows: in Section \cref{sec:background} we introduce the background to the problem and define our research questions. Section  \cref{sec:method} describes the method used. In Sections \cref{sec:results} and \cref{sec:discussion} we present the results and discuss their implications and limitations. Section \cref{sec:conclusions} presents conclusions and future research directions.

\section{Background}\label{sec:background}
\subsection{Global Software Development}

Global software development (GSD) promises cost effectiveness, access to large multi-skilled workforces, reduced time to market, and the opportunity to follow critical-path tasks around the clock \cite{beecham2015assessing}. These deciding factors, among others, have made GSD a daily reality in today's IT organizations, even though development environments are more complex and exhibit several challenges: physical distance, multiple time zones, the loss of `teamness,' culture differences, strategic issues, process differences, knowledge management and technical challenges \cite{beecham2015motivates,marinho2018globalcultural}.

In  particular,  GSD  is  normally  characterized  by stakeholders  from  different national  and  organizational cultures, located  in  separate  geographic  locations  and  time  zones,  using different  information  and  communication  technologies  to collaborate. Such  conditions  usually  result  in major problems in relation  to team  communication, coordination  and  collaboration \cite{niazi2016challenges}. Furthermore, key project-specific, team-specific, and distance-specific contextual factors may also impact team effectiveness. These include, for example, the nature of the contract, the application domain, the volatility of requirements, project personnel, site distribution, team  experience, and temporal, geographical, and socio-cultural distance between partners and sites \cite{Richardson_2012_Process}.

Traditionally, GSD follows a plan-driven, structured, waterfall approach, where tasks are allocated according to where they appear in the software development lifecycle \cite{estler2014agile,ylikotila2011collaboration}. By contrast, agile practices are considered capable of mitigating the challenges faced by GSD, as reported by several studies \cite{ramesh2006can,paasivaara2009using,hossain2009using,beecham2014using,noll2016agile}. However, Akbar and Safdar \cite{akbar2015short} find that the complexity of GSD and need for agility meant that organizations transitioning from co-located to distributed development were more inclined to adopt and tailor agile methods.

\subsection{Agile and Traditional Development Approaches}

Agile software development refers to a set of iterative and incremental software engineering methods that advocate an `agile philosophy' captured in the  Agile Manifesto \cite{fowler2001agile}. While mostly repackaging and re-branding previously well-known and appreciated software development practices, the agile movement can be considered an alternative to traditional software development methods \cite{conboy2009agility}. Agile development methods were believed to best suit small, co-located teams, and their success with small teams has inspired their use in large-scale software development \cite{noll2016agile} as reflected by the many scaled agile frameworks that have emerged over recent years \cite{ebert2017scaling}. However, fundamental assumptions regarding agile development are challenged when applying the methods at a large scale \cite{paasivaara2017adopting}.  

Agile methods were designed to accept and efficiently manage change \cite{cockburn2001agile}. It has been shown that agile methods have improved the satisfaction of both customers and developers. However, there is evidence that agile methods may not be a good fit for large undertakings \cite{dyba2009we}, as they require more coordination and heavier methodologies than smaller projects \cite{aitken_2013_comparative}.

Nerur et al. \cite{nerur2005challenges} describe the fundamental premise behind traditional methods as software that is wholly specifiable and should be developed with precise and comprehensive planning. Agile methods, on the other hand, consider that software can be built through on-going planning, improvement, and testing, based on fast feedback and change. Learning and adaptation should be embraced \cite{conboy2009agility}. Adapting the approach to the context will require balance in some areas \cite{vinekar2006can,akbar2015short}.

In plan-driven development, the architecture is defined before implementation and testing, whereas the architectural design appears as a result of on-going clarification in purist agile development. Traditionally, to construct a large software system developed by many teams, it is pivotal for the architecture to be agreed upon and communicated, adding to the bureaucracy and overhead associated with traditional methods. 
Indeed, GSD calls for different practices when it comes to architectural design, even when applying agile methods \cite{sievi2019challenges}.

\subsection{Hybrid Development Approaches}

Kuhrmann et. al. \cite{kuhrmann2017hybrid} proposed the following definition: \emph{`A hybrid software development approach is any combination of agile and traditional (plan-driven or rich) approaches that as organizational unit adopts and customizes to its own context needs (e.g., application domain, culture, process, project, organizational structure, techniques, technologies, etc.)'}

Some researchers argue that combining a mixture of traditional and agile practices is the best way to manage a project \cite{boehm2005management}. Boehm and Turner \cite{boehm2003using} mention that a changing world needs agile and plan-driven development methods. They characterize `house plots', where approaches are most likely to succeed, and they identify five critical dimensions, one being that a balanced strategy should be established to achieve a successful combination of agile and planned approaches.

Diebold and Zehler \cite{diebold2016right} examine the process of combining agile and traditional software development methods. They distinguish between revolutionary and evolutionary approaches, which differ in the methods’ order of occurrence. The authors defend the coexistence of both methods in software development projects.

Dingsoyr et al. \cite{dingsoyr2018key} summarize 12 pieces of advice they collected from 12 Scrum teams, stating that methods need to be adapted to needs during the program lifecycle. They argue that, while there is much good advice in frameworks such as the Scaled Agile Framework and Large-Scale Scrum, frameworks eventually limit organizations. It is thus preferable to combine agile and traditional approaches in global projects.

West et al. \cite{theocharis2015water} coined the term  `Water-Scrum-Fall' and hypothesized that hybrid development methods would become standard. They also provided evidence that development approaches are indeed used in combination (e.g. agile and traditional). Previous research on the use of development methods suggests that methods need to adapt to the work context \cite{fitzgerald2006customising}.

\section{Method}\label{sec:method}
The present study is part of a series of inquiries that have been conducted on data from a large international study on the use of hybrid development approaches, called \Helena: `(H)ybrid d(E)ve(L)opm(EN)t (A)pproaches in software systems development.’ The overall research project employs a mixed research method, with elements of both quantitative and qualitative research \cite{creswell2017research}. The project aims to provide a strong, empirically based assessment of the current state-of-practice in software and systems development \cite{kuhrmann2017hybrid}. 

\Helena was designed as an international research endeavour. The first stage prepared the data collection and tested the survey instruments. The second stage involved mass data collection conducted by an international consortium that comprised more than 60 partners from over 30 countries. More details can be found on the official website\footnote{\Helena Survey: https://helenastudy.wordpress.com/}. The results of the second stage were published at the 3rd \Helena Workshop. Further, all the authors of this article are involved in the stages of the study. The second stage results will fuel the third stage of the project by focusing follow-up in-depth research pertaining to outcomes of the second stage.

In this paper we use a subset of this data. The \Helena survey asked questions on how software organizations develop their software and what processes are used \cite{kuhrmann2017hybrid}. All published studies \cite{kuhrmann2019walking,klunder2019catching,tell2019hybrid,klunder2017helena,felderer2017hybrid,paez2017helena,scott2017initial} share the same general theme and use the same data sample, although they have different scope, variables, and designs. Our study is concerned with hybrid approaches in the context of GSD. A section of the survey was designed to shed light on current approaches to GSD in two different situations: what development approaches to GSD projects are used in practice, and which different development approaches are combined in practice.

Within the dataset of 691 completed responses\footnote{In total, the survey yielded 1,467 answers of which 691 participants completed the questionnaire \cite{helenastage2}.} \cite{helenastage2}, there were 263 companies that answered either option 3 or option 4 to the question: `Is the project or product your answer refers to carried out in a (globally) distributed manner?' with options: 

\begin{itemize}
    \item 1 = No;
    \item 2 = Yes, nationally (same country);
    \item 3 = Yes, regionally (same continent);
    \item 4 = Yes, globally
\end{itemize}

\subsection{Research Questions}

In this paper we focus solely on those 263 globally distributed organizations, and we defined the research questions listed in Table \ref{m:table5}.


\begin{table}
\scriptsize
\centering
\caption{Research Questions Overview}
\label{m:table5}
\begin{tabular}{@{}ll@{}}
\toprule
\multicolumn{2}{l}{Research Questions} \\ 
\midrule
RQ1 & What development approaches do GSD projects use?  \\
RQ2 & How do GSD projects combine development approaches (if at all)?  \\ 
RQ3 & Do combined (`hybrid') development approaches tend toward agile \\
 & or traditional development? \\ 
\bottomrule
\end{tabular}
\end{table}

\subsection{Survey Instrument}

\Helena used an online survey \cite{ciolkowski2003practical, shull2007guide} to collect data from practitioners about the development approaches they use in their projects.

\subsubsection{Instrument Development}

The \Helena research  team used a multi-stage approach to develop the survey instrument. 
Initially, three researchers developed the questionnaire and tested it with 15 German practitioners to evaluate the agreement. Based on the feedback, a team of eleven researchers from across Europe revised the survey and translated it to English. 
A first revised questionnaire public test (referred to as ``Stage 1'') that included 25 questions was conducted in 2016 in Europe. 
This public test yielded 69 data points, which were analysed and used to initiate the next stage of the study. 

In Stage 2, the research team was expanded to 75 researchers from around the world. 
Also, the questionnaire was revised to improve structure, scope, relevance, question accuracy, value ranges for variables, and significance of the topics included. 
The revised questionnaire was translated and made available in English, German, Spanish, and Portuguese. Further details of the instrument are presented in \cite{helenastage2} )

\subsubsection{Instrument Structure}

The final survey consisted of 38 questions divided into five parts: Demographics (10 questions), Process Use (13 questions), Process Use and Standards (5 questions), Experiences (2 questions) and Closing (8 questions). 
However, because some questions may not be relevant depending on answers to previous questions, the survey responses do not always contain 38 answers \cite{helenastage2}.

\begin{table*}
\scriptsize
\centering
\caption{\Helena questionnaire list (conditional questions for the different paths are omitted in the table)}
\label{table4}
\begin{tabular}{@{}lclcc@{}}
\toprule
No. & \multicolumn{1}{l}{Group\footnotemark[1]} & Question & \multicolumn{1}{l}{Scale\footnotemark[2]} & \multicolumn{1}{l}{opt} \\ \midrule
1. & M & What is your company's size in equivalent full-time employees (FTEs)? & SC & 5 \\
2. & M & Do you participate in globally software project? & SC & 4 \\
3. & M & In which country are you personally located? & FT &  \\
4. & M & What is the major role you have in this project? & SC & 9 \\
5. & M & What is the size of the project? & SC & 5 \\
6. & M & How many years of experience do you have in software and systems development? & SC & 5 \\ 
7. & PU & Do you combine different development approaches? & YN &  \\
8. & PU & \begin{tabular}[c]{@{}l@{}}For the following standard activities in the project or product development, please indicate\\  to which degree you carry out project activities in a more traditional or more agile manner.\end{tabular} & LI &  \\
9. & PU & How were the combinations of development approaches in your company developed? & MC & 3 \\
10. & PU & How was your project-specific development approach defined? & MC & 6 \\
11. & PU & Which of the following frameworks and methods do you use? & RT & 7 \\
12. & PU & What are the overall goals that you aim to address with your selection and combination of development approaches? & MC & 2 \\
13. & PU & To what degree did the combination of approaches help you to achieve your goals? & RT & 10  \\ \bottomrule
\end{tabular}
\\ \footnotemark[1]\emph{Legend for groups: M=metadata, PU=process use, E=experience.}
\\ \footnotemark[2]\emph{Legend for scales: YN=yes/no, SC=single choice, MC=multiple choice/select, LI=5-item Likert scale, RT= Rating: 7 point scale.}
\end{table*}

\subsection{Data Collection Procedure}

The survey was promoted through personal contacts of the participating researchers, through posters at conferences, and through posts to mailing lists, social media channels \cite{robson2016real}. The target population was members of IT clusters and networks, and we used LinkedIn, Twitter, Facebook, Xing, and ResearchGate to promote the survey within the relevant communities. A web survey with different kinds of scale questionnaires and demographic information collection was distributed to the target population.

Potential biasing factors in the results include those that are common to self-selected written surveys with convenience sampling: response bias (i.e. people who responded may have different characteristics from those who did not), coverage errors (i.e. the representation of participants may not be balanced across different sub-communities), and item bias (i.e. some questions may have been skipped intentionally or unintentionally).

\subsection{Data Analysis Procedure}

We utilised several methods to provide answers to the research questions. For all research questions, we used descriptive statistics, to provide tables and charts for process use and selection.

To provide a context we analysed the types of companies involved (size, how many sites, what countries are represented, and roles of respondents). We also analysed (a) what types of development are used in GSD and (b) how the different development approaches are combined. 

To further analyze the result set, we used the three categories: \emph{traditional}, \emph{agile}, and \emph{generic}, based on the definitions provided by Kurhman and colleagues~\cite{kuhrmann2017hybrid}. 
We classified the following approaches as agile: Scrum, Safe, Lean, LESS, Nexus, XP, Kanban, DevOps, ScrumBam, Crystal, DSDM and Feature-driven development (FDD).  
Similarly, we classified the following approaches as traditional: Waterfall, Spiral Model, V-Model, RUP, PRINCE2 and SSADM. 
Other approaches -- Iterative development, Domain-Drive Design (DDD), Model Driven Architecture (MDA), Team Software Process (TSP), Personal Software Process (PSP) -- were classified as generic, since the approach does not fit into either the agile or traditional category. 
While Iterative development is a key feature of agile development approaches~\cite{larman2003iterative}, some traditional approaches (RUP and the Spiral Model, for example) also incorporate iterations~\cite{larman2003iterative}; consequently, we followed Kuhrmann et al. \cite{kuhrmann2017hybrid} and classified Iterative development in the generic category.

Many projects used combinations of methods in different categories.
In light of this, we assigned such projects to a category according to the following rules:

\begin{itemize}
    \item Agile + Generic = Agile;
    \item Traditional + Generic = Traditional;
    \item Agile + Traditional = Hybrid.
\end{itemize}

These rules are based on the observation that ``generic'' processes could be either agile or traditional depending on context, which in this case is determined by the other methods used, which are either agile or traditional.
Each organization was categorised based on the approaches they stated they used when answering the survey.

The thirteen questions from the \Helena survey which we analysed are listed in Table \ref{table4}. 
If a company stated that it combined different development approaches (question 7), we then looked at questions 8, 10 and 12 to confirm whether this was indeed the case.
Of the 263 global companies, 204 selected 'Yes' to question 7.  
In checking the other answers from these 204 respondents, we were able to establish that companies used Hybrid approaches (Traditional + Agile) in 189 projects that were detailed in the survey (see Table \ref{r:table_projects}). Additionally, through undertaking this exercise, we verified that the answer given to question 7 is due to how the respondents interpreted `combination'. Many respondents indicated the use of multiple agile or traditional methods; it is possible that respondents interpret this as `combining' methods, even when they are in the same category.    

\begin{table}
\scriptsize
\centering
\caption{Number of projects by category.}
\label{r:table_projects}
\begin{tabular}{@{}lll@{}}
\toprule
\textbf{Class} & \textbf{Num} & \textbf{\%}   \\ \midrule
Agile          & 65           & 25.0           \\
Traditional    & 5            & 2.0            \\
Hybrid         & 189          & 72.0           \\
Other          & 4            & 1.0            \\ \midrule
\textbf{Total} & \textbf{263} & \textbf{100} \\ \bottomrule
\end{tabular}
\end{table}

\section{Results}\label{sec:results}
In this section, we summarize the results of our investigation. We first show descriptive statistics describing the demographics of the participants. We then proceed to data analysis related to our research questions (\ref{m:table5}).

\subsection{Study Population}

The survey yielded 263 valid and complete responses related to GSD projects,
from 33 different countries, covering 5 continents, as shown in Table \ref{r:table_continent}. 
These projects were situated in 18 micro-sized, 29 small, 55 medium, 61 large, and 100 very large companies.

\begin{table}
\scriptsize
\centering
\caption{Project location (\emph{n=263}, from 33 countries)}
\label{r:table_continent}
\begin{tabular}{@{}ll@{}}
\toprule
\multicolumn{1}{c}{\textbf{Continent/Country}} & \multicolumn{1}{c}{\textbf{Number}} \\ \midrule
\textbf{Africa (2)}                            &                                     \\
Algeria                                        & 1                                   \\
Uganda                                         & 1                                   \\
\textbf{America (85)}                          &                                     \\
Argentina                                      & 28                                  \\
Brazil                                         & 6                                   \\
Canada                                         & 4                                   \\
Chile                                          & 2                                   \\
Colombia                                       & 1                                   \\
Costa Rica                                     & 25                                  \\
United States                                  & 18                                  \\
Uruguay                                        & 1                                   \\
\textbf{Asia (24)}                             &                                     \\
Armenia                                        & 1                                   \\
China                                          & 4                                   \\
Christmas Island                               & 1                                   \\
India                                          & 9                                   \\
Japan                                          & 1                                   \\
Pakistan                                       & 4                                   \\
Saudi Arabia                                   & 1                                   \\
Turkey                                         & 3                                   \\
\textbf{Europe (141)}                          &                                     \\
Austria                                        & 10                                  \\
Denmark                                        & 17                                  \\
Estonia                                        & 4                                   \\
Finland                                        & 4                                   \\
Germany                                        & 49                                  \\
Ireland                                        & 10                                  \\
Italy                                          & 1                                   \\
Poland                                         & 1                                   \\
Portugal                                       & 4                                   \\
Russia                                         & 3                                   \\
Spain                                          & 14                                  \\
Sweden                                         & 8                                   \\
Switzerland                                    & 13                                  \\
United Kingdom                                 & 3                                   \\
\textbf{Oceania (11)}                          &                                     \\
New Zealand                                    & 11                                  \\ \bottomrule
\end{tabular}
\end{table}

The size of the projects were characterized as small (7), medium (21), large (37), and very large (198). 
The respondents had a range of experience in the area of software and systems development, ranging from $<1$ year (8), through 1--2 years (12), 3--5 years (35), 6--10 years (49), to $>$10 years (159). 
Twenty-three percent of respondents were Developers, 22\% Project/Team Managers, 11\% Product Manager/Owners, 10\% Architects, 8\% Quality Managers 6\% Scrum Master/Agile Coaches, 6\% C-level Management (e.g. CIO, CTO), 3\% Testers, 3\% Analyst/Requirements Engineers, 1\% Trainers and 8\% other role.

\subsection{Development Approaches used in GSD}

\begin{figure}[!h]
    \centering
    \includegraphics[width=0.5 \textwidth]{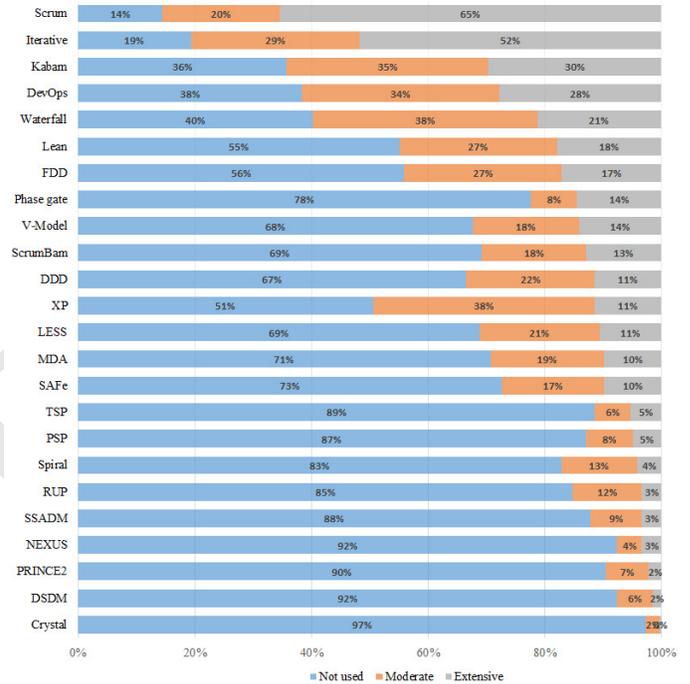} 
    \caption{Frameworks and methods used by GSD projects (\emph{n=263}).}
  \label{fig:frameworks-methods}
\end{figure}

\cref{fig:frameworks-methods} presents the frameworks researched and the degree of utilization by the global projects. We collapsed the responses to simplify the representation of the framework usage:

\begin{itemize}
    \item Not used = we do not know or we do not know if we use it or we never use it;
    \item Moderate = we rarely use it or we sometimes use it;
    \item Extensive = we often use it or we always use it;
\end{itemize}

The participants state that, among agile approaches, the most frequently used framework is Scrum, followed by Kanban. 
Iterative development was the most used by the GSD projects that responded to this study. However, a waterfall approach was the most commonly used traditional framework.  

\subsection{Combination of Development Approaches}

Respondents were asked how framework combinations and practices in the participants' organization were developed. 
83\% reported that they evolved from past projects over time, and 43\% reported that they were planned as part of a process-improvement program  (\cref{tab:how-combinations-developed}). 

\begin{table}[h]
\scriptsize
\centering
\caption{How were the combinations of development frameworks and methods in your company developed? (\emph{(Of the 189 participants, 154 answered these affirmations}) }
\label{tab:how-combinations-developed}
\begin{tabular}{@{}lll@{}}
\toprule
Option                                           & \#  & \%   \\ \midrule
Planned as part of a process improvement program & 69  & 43.0 \\
Evolved as learning from past projects over time & 128 & 83.0 \\ \bottomrule
\end{tabular}
\end{table}

Participants were asked to define their project-specific development approach (see \cref{tab:process-selection-tailoring} for results).  
25\% of projects reported  project-specific process selection and tailoring followed defined rules, while 24\% of projects selected  specific practices and methods  according to customer demands. 

\begin{table}[h]
\scriptsize
\centering
\caption{Overview of the actual process selection and tailoring in particular projects. (\emph{n=189}).}
\label{tab:process-selection-tailoring}
\begin{tabular}{@{}lll@{}}
\toprule
Option                                           & \#  & \%   \\ \midrule

Project specific process selection and tailoring follows defined rules & 48 & 25 \\
Specific practices and methods are selected according to customer demands & 45 & 24 \\
A project manager tailors the process in the beginning of a project & 30 & 16 \\
Specific practices and methods are selected in the project on demand & 30 & 16 \\
Project specific process selection and tailoring is supported by tools & 13 & 7 \\
The process is not tailored at all & 10 & 5 \\
Other & 13 & 7 \\ \bottomrule
\end{tabular}
\end{table}

Participants were asked to list the overall goals that drive their selection of development approaches. 
The main objectives reported were: improved productivity, improved planning and estimation, improved external product quality and improved frequency of delivery to customers. 
\cref{tab:combination-goals} shows that global hybrid project delivery is merely a combination of selected approaches working together in order to increase performance beyond what any single methodology can achieve. The precise combination of techniques varies from one project to another.

\begin{table}
\scriptsize
\centering
\caption{Overall goals when selecting and combining approaches (\emph{n=189}).}
\label{tab:combination-goals}
\begin{tabular}{@{}ll@{}}
\toprule
\textit{\textbf{Goal: Improved...}}                            & \textbf{\%} \\ \midrule
Productivity                                                   & 67.0        \\
Planning and estimation                                        & 65.0        \\
External product quality                                       & 64.0        \\
Frequency of delivery to customers                             & 64.0        \\
Adaptability and flexibility of the process to react to change & 58.0        \\
Time to market                                                 & 58.0        \\
Project monitoring and controlling                             & 53.0        \\
Client involvement                                             & 49.0        \\
Internal artifact quality                                      & 48.0        \\
Employee satisfaction                                          & 43.0        \\
Knowledge transfer and learning                                & 41.0        \\
Risk management                                                & 39.0        \\
Return-on-investment cycles                                    & 38.0        \\
Reuse for project artifacts                                    & 35.0        \\
Maturity of the company                                        & 31.0        \\
Ability of the company to develop critical systems             & 28.0        \\
Tool support                                                   & 28.0        \\
Staff education and development                                & 26.0        \\ \bottomrule
\end{tabular}
\end{table}

The majority of participants answered that a hybrid development approach appropriately supports business projects (e.g. concerning internal or customer goals). Further, this approach helps to achieve good product quality. Participants said that their current development approach adequately addressed the external requirements (e.g. standards). The approach supported the work and helped to hasten its time to market. However, the majority of participants noted that they would change or improve their current development approach. We believe this is an example of continuing quality improvement through hybrid designs.

On the other hand, 43\% of the participants stated that the combination of methods and frameworks improves the satisfaction of the team. According to \cite{klunder2017fake}, using agile methods and practices exclusively does not necessarily lead to a more competent development team. Rather, doing so can undermine the quality and productivity of software, as well as social factors such as team motivation and performance. Thus, we believe that, in addition to combining methods for better product results, it is necessary to structure a philosophy for hybrid teams which can improve team satisfaction.

\begin{figure}[!h]
    \centering
    \includegraphics[width=0.5 \textwidth]{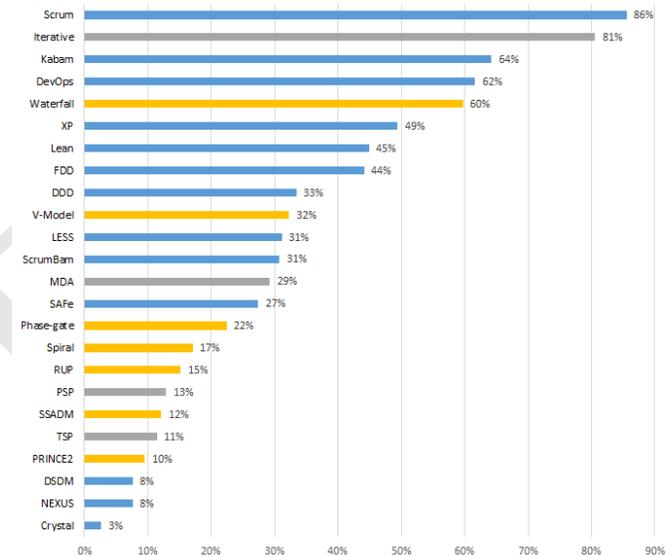} 
    \caption{Framework usage frequency (\emph{n=263}). colors: agile: blue, traditional: orange, generic: grey.}
  \label{fig:framework-usage-freq}
\end{figure}

\cref{fig:framework-usage-freq} shows a classification of the different approaches where we present data from the 263 companies who are involved in GSD. For instance, a project might adopt Scrum at all times, but use some plan-driven practices. Thus, in the sequence, global projects adopt more scrum, iterative development, and kanban-based approaches. Among the frameworks that were created to scale agility, the use of LESS and Safe was followed by Nexus.

We performed a comparison of how frequently traditional and agile methods are combined in GSD projects (\cref{tab:combinations-table-trad}); 
the rows show agile-specific approaches and the columns show the frequency (extensive or moderate) of each traditional approach adopted in the GSD projects. 
For example, 89\% of GSD projects always adopt Scrum. Combinations are adopted when Scrum is used, and the frequency of adoption of waterfall approach was 18\% `extensive' and 41\% `moderate'.
This comparison indicates that GSD projects using an agile approach also use traditional approaches during execution.

\begin{table*}[ht]
\tiny
\centering
\caption{Frequency with which traditional methods are performed when Agile practices are at least \emph{rarely} performed by GSD projects.}
\label{tab:combinations-table-trad}
\begin{tabular}{@{}lcccccccccccccccl@{}}
\toprule
\textit{\textbf{}} & \multicolumn{2}{c}{PRINCE2 (25\protect\footnotemark[2]) 10\%} & \multicolumn{2}{c}{RUP (40) 15\%} & \multicolumn{2}{c}{Spiral (45) 17\%} & \multicolumn{2}{c}{Waterfall (157) 60\%} & \multicolumn{2}{c}{Stage Gate (59) 22\%} & \multicolumn{2}{c}{SSADM (32) 12\%} & \multicolumn{2}{c}{V-model (85) 32\%} & \multicolumn{2}{c}{Neither trad.} \\ 
Agile practice     & extensive\protect\footnotemark[1] & moderate         & extensive      & moderate       & extensive       & moderate         & extensive         & moderate          & extensive          & moderate          & extensive       & moderate        & extensive        & moderate        & \multicolumn{2}{c}{nor waterfall} \\ \midrule

Scrum (225) 86\%   & (6) 3\%      & (18) 8\%     & (8) 4\%    & (29) 13\%  & (8) 4\%     & (32) 14\%    & (41) 18\%     & (93) 41\%     & (34) 15\%      & (18) 8\%      & (5) 2\%     & (23) 10\%   & (31) 14\%    & (43) 19\%    & \multicolumn{2}{c}{(56) 25\%}     \\
Kanban (169) 64\%  & (5) 3\%      & (12) 7\%     & (5) 3\%    & (22) 13\%  & (5) 3\%     & (21) 12\%    & (32) 19\%     & (67) 40\%     & (29) 17\%      & (15) 9\%      & (4) 2\%     & (13) 8\%    & (28) 17\%    & (34) 20\%    & \multicolumn{2}{c}{(40) 24\%}     \\
DevOps (162) 62\%  & (5) 3\%      & (15) 9\%     & (6) 4\%    & (18) 11\%  & (2) 1\%     & (23) 14\%    & (29) 18\%     & (62) 38\%     & (26) 16\%      & (14) 9\%      & (1) 1\%     & (16) 10\%   & (15) 9\%     & (31) 19\%    & \multicolumn{2}{c}{(43) 27\%}     \\
XP (130) 49\%      & (3) 2\%      & (11) 8\%     & (8) 6\%    & (23) 18\%  & (3) 2\%     & (27) 21\%    & (22) 17\%     & (60) 46\%     & (13) 10\%      & (13) 10\%     & (2) 2\%     & (14) 11\%   & (12) 9\%     & (32) 25\%    & \multicolumn{2}{c}{(27) 21\%}     \\
Lean (118) 45\%    & (5) 4\%      & (11) 9\%     & (7) 6\%    & (18) 15\%  & (7) 6\%     & (21) 18\%    & (20) 17\%     & (53) 45\%     & (24) 20\%      & (14) 12\%     & (6) 5\%     & (12) 10\%   & (24) 20\%    & (26) 22\%    & \multicolumn{2}{c}{(22) 19\%}     \\
FDD (116) 44\%     & (3) 3\%      & (12) 10\%    & (4) 3\%    & (16) 14\%  & (8) 7\%     & (18) 16\%    & (26) 22\%     & (50) 43\%     & (22) 19\%      & (13) 11\%     & (4) 3\%     & (15) 13\%   & (23) 20\%    & (24) 21\%    & \multicolumn{2}{c}{(20) 17\%}     \\
LESS (82) 31\%     & (3) 4\%      & (8) 10\%     & (6) 7\%    & (15) 18\%  & (3) 4\%     & (14) 17\%    & (15) 18\%     & (39) 48\%     & (15) 18\%      & (10) 12\%     & (4) 5\%     & (9) 11\%    & (11) 13\%    & (21) 26\%    & \multicolumn{2}{c}{(12) 15\%}     \\
Scrumban (81) 31\% & (3) 4\%      & (8) 10\%     & (5) 6\%    & (14) 17\%  & (3) 4\%     & (15) 19\%    & (18) 22\%     & (33) 41\%     & (12) 15\%      & (10) 12\%     & (3) 4\%     & (13) 16\%   & (12) 15\%    & (23) 28\%    & \multicolumn{2}{c}{(14) 17\%}     \\ 
SAFe (72) 27\%     & (3) 4\%      & (11) 15\%    & (7) 10\%   & (16) 22\%  & (4) 6\%     & (15) 21\%    & (16) 22\%     & (37) 51\%     & (18) 25\%      & (9) 13\%      & (3) 4\%     & (13) 18\%   & (13) 18\%    & (18) 25\%    & \multicolumn{2}{c}{(7) 10\%}      \\
DSDM (20) 8\%      & (1) 5\%      & (4) 20\%     & (4) 20\%   & (6) 30\%   & (3) 15\%    & (6) 30\%     & (4) 20\%      & (11) 55\%     & (4) 20\%       & (2) 10\%      & (4) 20\%    & (5) 25\%    & (2) 10\%     & (6) 30\%     & \multicolumn{2}{c}{(3) 15\%}      \\
Nexus (20) 8\%     & (2) 10\%     & (6) 30\%     & (2) 10\%   & (8) 40\%   & (1) 5\%     & (6) 30\%     & (3) 15\%      & (8) 40\%      & (3) 15\%       & (3) 15\%      & (1) 5\%     & (8) 40\%    & (2) 10\%     & (9) 45\%     & \multicolumn{2}{c}{(0) 0\%}       \\
Crystal (7) 3\%    & (1) 14\%     & (3) 43\%     & (2) 29\%   & (3) 43\%   & (1) 14\%    & (3) 43\%     & (1) 14\%      & (3) 43\%      & (1) 14\%       & (3) 43\%      & (1) 14\%    & (2) 29\%    & (2) 29\%     & (3) 43\%     & \multicolumn{2}{c}{(0) 0\%}       \\ \hline
Agile: (254) 96\%  & (6) 2\%      & (19) 7\%     & (9) 4\%    & (31) 12\%  & (11) 4\%    & (34) 13\%    & (56) 22\%     & (101) 40\%    & (38) 15\%      & (21) 8\%      & (9) 4\%     & (23) 9\%    & (37) 15\%    & (48) 19\%    & \multicolumn{2}{c}{(66) 26\%}     \\ \bottomrule
\end{tabular}
\\ \footnotemark[1]\emph{The 'moderate' column indicates the method was used rarely or sometimes, while the 'extensive' column means a method was used often or always.}
\\ \footnotemark[2] \emph{The number given in parentheses indicates the number of respondents using the approach.}
\end{table*}

We found that, in the combination of approaches in an `extensive' way, the Waterfall approach is the most combined with agile (22\%), followed by Stage-Gate model and V-model (15\% each). Further information regarding GSD projects can be derived from this table: when Scrum is used, Prince2 is the least-used approach; the most combined approaches with Crystal were RUP and the V-model; when DSDM is used, it is most often combined with the Waterfall, RUP, V-model and Stage Gate traditional approaches.

\begin{table}
\tiny
\centering
\caption{Comparison of how frequent traditional and agile methods are combined.\protect\footnotemark[1]}
\label{summary-table}
\begin{tabular}{l|ccccc|c}
\toprule
 \multicolumn{6}{c}{Traditional} 
\\
Agile             &            never&           rarely&        sometimes&            often&           always&            TOTAL\\
\midrule
never             &        (4)  44\%&        (0)   0\%&        (0)   0\%&        (3)  33\%&        (2)  22\%&        (9)   3\%\\
rarely            &        (1)  14\%&        (0)   0\%&        (1)  14\%&        (2)  29\%&        (3)  43\%&        (7)   3\%\\
sometimes         &        (6)  19\%&        (2)   6\%&        (8)  25\%&        (7)  22\%&        (9)  28\%&       (32)  12\%\\
often             &       (17)  16\%&       (22)  21\%&       (23)  21\%&       (36)  34\%&        (9)   8\%&      (107)  41\%\\
always            &       (41)  38\%&       (19)  18\%&       (19)  18\%&       (13)  12\%&       (16)  15\%&      (108)  41\%\\
\midrule
TOTAL             &       (69)  26\%&       (43)  16\%&       (51)  19\%&       (61)  23\%&       (39)  15\%&      (263) 100\%\\
\bottomrule
\end{tabular}
\\ \footnotemark[1]\emph{The TOTAL column indicates the total frequency with which projects perform agile methods at the level shown; the TOTAL row indicates the total frequency with which project perform traditional methods a the level shown}
\end{table}

Table \ref{summary-table} shows that combining agile and traditional methods is common. For example, looking at the columns and rows (rarely, sometimes, often and always), we see that the combination of methods occurs in 72\% of the projects, as shown in Table \ref{r:table_projects}. Our results strongly indicate that it is easier to find hybrid GSD projects than purely agile ones. It is understood that opportunities to adapt and customize each of these approaches must meet the specific needs of the current situation to allow organizations the flexibility to develop a project and use the best methods in a particular aspect of the work.

\subsection{Agility of Hybrid Approaches}

\begin{figure}[!h]
    \centering
    \includegraphics[width=0.4 \textwidth] {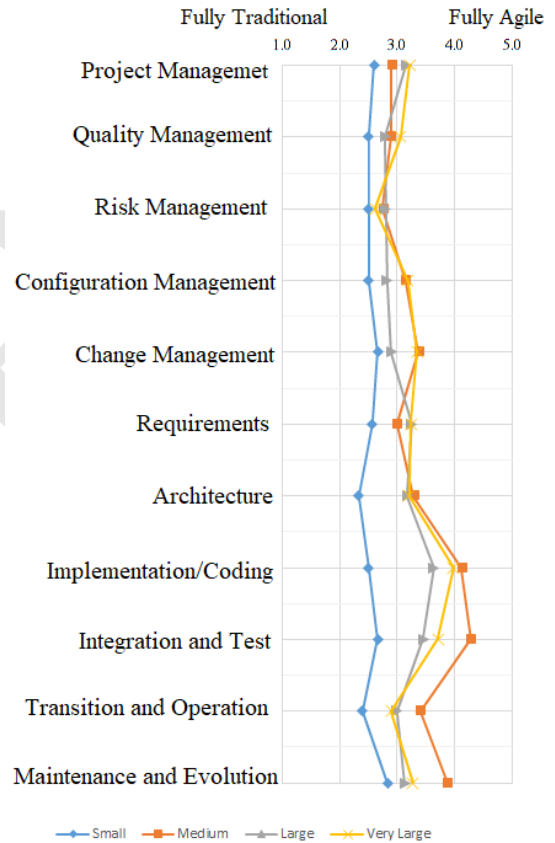} 
    \caption{Activities implementation in GSD projects ( 1=fully traditional to 5=fully agile; Average rating of \emph{n=189} hybrid projects.}
  \label{fig:swebok}
\end{figure}

In the survey, participants were asked, ``For the following standard activities in the project or product development, please indicate to which degree you carry out activities in a more traditional or more agile manner''; 
participants rated their approach to performing each of the disciplines on a five point scale comprising ``Fully Traditional'' (1), ``Mainly Traditional'' (2), ``Balanced between Traditional and Agile'' (3), ``Mainly Agile'' (4) and ``Fully Agile'' (5).  
\cref{fig:swebok} summarizes responses using the average rating among small, medium, large, and very large companies.

In general, the responses indicate that activities are implemented in a balanced way between agile and traditional methods.
However, management activities such as Risk Management and Quality Management are more traditionally oriented, whereas development activities, such as Implementation/Coding and Integration/Testing, tend to be conducted in a more agile manner. 
Also,  \cref{fig:swebok} shows that small companies tend to use traditional methods more than medium, large and very large companies, that tend to be more agile. 

\section{Discussion}\label{sec:discussion}
Agile adoption is commonplace, and as shown in the GSD literature, operating in a distributed setting is not a barrier to adopting agile methods \cite{vallon2018systematic}. 
In this study, however, we found that, in practice, hybrid approaches are adopted in the majority (72\%) of GSD projects, while 25\% only use agile methods, and 5\% use only traditional methods.

Our study consisted of responses from 263 participants in GSD projects, of whom 198 participated in very large global projects and 61\% had more than 10 years' experience.  
Regarding research question RQ1 (\cref{fig:framework-usage-freq}), participants reported the most adopted approaches are, from highest to lowest: Scrum, Iterative Development, Kanban, DevOps and Waterfall. 
When combining approaches (RQ2 \cref{tab:combinations-table-trad}) at different times in the project, the responses indicate that projects often use Scrum and Waterfall; and out of 263 GSD projects, 189 used hybrid software development approaches.

Most of the GSD projects represented in the survey results are neither predominantly planned or agile in their approach.
Rather, the approaches appear to be  adjusted to reflect the challenges and opportunities of each initiative (\cref{tab:process-selection-tailoring}). 
For example, 
\cref{fig:framework-usage-freq} shows that 86\% of GSD projects adopt Scrum, and \cref{tab:combinations-table-trad} shows that, when adopting Scrum, there is a requirement to combine it with traditional approaches. This result agrees with the conclusion of Lous et al. \cite{lous2017scrum} who performed a systematic review of the Scrum literature with GSD and came to the conclusion that all Scrum adopters need to (extensively) adjust the core assets of Scrum to a GSD context. 
Likewise, Fitzgerald and Stol \cite{fitzgerald2018future} affirm that GSD requires some tailoring of agile methods that were not intended for such settings. While agile methods have moved well beyond this, with frequent use of agile methods in projects within globally distributed teams, many challenges remain, such as those pertaining to cultural differences, breakdowns in communication, and optimum practices for distributing development work across sites. Individual successes rely heavily on the particular organizational context, which is increasingly recognized as an essential factor \cite{dybaa2012works}.

Bass \cite{bass2016artefacts} argued that several agile artefacts, such as user stories and sprint backlogs, are used in global projects. Further, teams adopt elements of an agile culture. However, to deal with release plans, test plans, and product architecture implementation, the methodology adopted follows a plan-driven approach. Bass \cite{bass2016artefacts} states that teams skilfully blend agile artefacts with traditional plan-based artefacts because of the need for coordination, and that this combination is reflected in the quality.
Finally, Richter et al. \cite{richter2016problems} analysed agile GSD in an organization and identified several issues. They concluded that there is no single solution or universal software development approach. Teams and organizations use different strategies to address the manifold of challenges in software development projects.
As a consequence, it appears that there is a need for agile practices, which were originally designed for collocated teams, to be modified to work in distributed environments. Jalali and Wohlin \cite{jalali2012global} also found this to be necessary. 

Diebold et al. \cite{diebold2015practitioners} showed that it is common to use a textbook method, but one that adapts to specific contexts (the project's or the company's). This was reflected in GSD projects in the results of \cref{tab:process-selection-tailoring}, which  shows an overview of the actual process selection and tailoring in particular projects.

An agile structure requires engagement based on the interaction of all teams, whereas traditional models determine the specific phases in which team activities occur \cite{theocharis2015water}. In large and distributed projects, the main focus should be to keep the organizational structure small. The hierarchical structure should be as flat as possible, with more teams and fewer levels of management. Teams can be real or virtual, but the main difference in similar groups formed when following traditional methods of project management is that teams must have cross-links. One member of each team must attend the major meetings of the other teams. Having interconnected, self-organized teams and a flat organizational structure allows for agile teams \cite{dikert2016challenges}.

\cref{tab:combinations-table-trad} presents combinations of the use of traditional approaches. Of the surveyed projects that used scaling agile frameworks, such as LESS, SAFe and Nexus, the majority used traditional methods moderately or extensively.
In scaled projects, the existence of an architecture team is significant, as it guides all teams on the proper implementation of the requirements, reducing development costs by proactively analysing alternatives and finding solutions that best meet the needs, adaptability, and future use \cite{leffingwell2007scaling}. 
As shown by \cite{putta2018benefits}, several case studies show that moving to SAFe feels like moving back to plan-driven methods (such as Waterfall and RUP) which include fixed increments, centralized planning, a loss of incremental and iterative development, and too much detail. Although LESS applies agile practices, we also investigated agile/plan-driven. Likewise, \cite{dikert2016challenges} suggests a way of combining plan-driven with agile. However, proponents of agility are unlikely to sanction such a mix.

To maintain control of large and distributed projects, the natural inclination is to introduce layers of management, improved policies, new processes, and checkpoints based on the project competencies of managers. For this reason, many of the practices and tools identified for distributed agile projects are the same as those followed in projects using traditional project management approaches. Examples include the need for an architecture team that positively influences both traditional and agile methodologies, along with quality assurance and integration teams that have similarly scaled functions for large, distributed teams \cite{sievi2019challenges}.

\subsection{Threats to Validity}

\subsubsection{Internal Validity} Each data transcript from the survey was faithfully represented in answers or comments in the study. Sampling of the participants was conducted with maximum variation, following a revised and approved questionnaire and synthesis tools, such as SPSS and Microsoft Excel, in order to increase the internal validity.

\subsubsection{External Validity} In this study, independent researchers were consistently involved. Furthermore, results were compared with previous studies to find a reference for data interpretation. However, in order to generalise the results, further research in more regions is necessary. Also, the survey did not include the distribution of the teams (we do not know the geographic distance, temporal distance, cultural distance nor team size). Our results are based on responses taken from projects that are geographically distributed (across continents and countries).

\section{Conclusions}\label{sec:conclusions}
Problems related to the complexity of software development and increased focus on coordination, communication, and collaboration are the main reasons for the interest in applying agile methods to GSD \cite{paasivaara2009using}. 
GSD uses asynchronous communication and computer-mediated activities, but the applicability of these coordination mechanisms is generally insufficient \cite{niazi2016challenges}. 
Our survey illustrates that most companies surveyed applied a combination of traditional and agile methods in their distributed projects.  
Although the agile approach has many advantages, our results suggest that it is not a universal solution. 
Consequently, we often see companies using a combination of agile and traditional methods. 
These hybrid approaches seem to be products of a trade-off between reaping the benefits of an iterative, flexible, and collaborative approach with maintaining some level of planning and structure.

In this study, using the publicly available \Helena Survey \cite{helenastage2} , we aimed to highlight how GSD projects are adopting hybrid approaches.
We looked at which frameworks are being used, and examine how GSD projects still use traditional plan-driven approaches. 
We analysed 189 responses from projects in which traditional and agile approaches were combined. 
Our results led us to the conclusion that most global software projects are neither purely agile or purely traditional in their approach to software development, but rather combine agile and traditional methods.
The most used and combined approaches in the responses were Scrum and Waterfall. 
In particular, we found that projects which adopted agile scaling frameworks such as SAFe or LESS, also employed plan-driven methods.
In summary, our study supports the findings of Aitken \& Ilango \cite{aitken_2013_comparative}, that ``there is nothing really incompatible with applying all the principles and values of agile software development, along with most (if not all) of the practices, to traditional software engineering''.

\subsection{Limitations}

Based on our data collection, we note four limitations.  First, the data does not represent all agile, plan-driven, and GSD specific approaches. Second, there is relatively low representation of United States-based projects in the sample population, the first user community to publicise the agile movement, and the most experienced and most populous among agile user communities globally. In this study, although the United States is the leading country, with a representation of 7\% of projects, it can still be considered underrepresented, since according to Agile Alliance\footnote{https://www.agilealliance.org/} 35\% of all agile user groups (29 of 83) are based in the United States. Further, their website claims that more than 58\% of Agile Alliance users are from the United States. However, we do have 35 different countries represented in our dataset.  Third, our data does not include information on team size, the extent of the distribution, nor the specific countries involved in the distributed project.  All of these may have an impact on why a certain development approach is taken.  Finally, the sample size is small, considering the large population of the agile and GSD community. More respondents could provide more robust and accurate statistical calculations and analyses, and could also include other methods that were not observed.

Notwithstanding these observations, what is obvious to us as researchers is that, as companies are continually combining agile, traditional and global software development methods, we have identified a need for an approach by which companies can build their own bespoke software development method.  


\subsection{Future Research}

Since our results show that hybrid methods are the current state of practice in \GSD projects, this should be taken as a new baseline for future research. The strategies applied today are still some way from perfect when it comes to devising hybrid methods. Therefore, the first main future direction for research is to provide better strategies to devise hybrid methods.

Also, we find that agile scaling approaches still use traditional process elements for specific purposes. Notably, the finding that the waterfall model, even criticised a lot, is still frequently used. So, studying the reasons for the agile scaling approaches uses more traditional process allows us to understand potential directions better to mitigate the use of hybrid methods.
Finally, we should develop strategies suitable for use by practitioners when being confronted with the need to integrate different processes in \GSD projects.

\textbf{Acknowledgements}
We thank all the study participants and the researchers involved in the \Helena project for their great effort in collecting data points. This work was supported, in part, by Science Foundation Ireland grant no. 13/RC/2094.


\footnotesize
\bibliographystyle{IEEEtran}
\bibliography{ms}



\end{document}